\documentstyle[psfig,amssym,aps,preprint]{revtex}
\newcommand{\Li}[1]{\mbox{Li}_2\left(#1\right)}
\newcommand{\bea}{\begin{eqnarray}}
\newcommand{\eea}{\end{eqnarray}}
\newcommand{\be}{\begin{equation}}
\newcommand{\ee}{\end{equation}}

\draft
\makeatletter

                \def\@preprint{}
                \def\preprint#1#2#3{%
                \ifpreprintsty
                \def\@preprint{
                \noindent \hbox{#1}\hfill\hbox{#2}\\
                \hbox{#3}\vskip -2ex}%
                \fi
                }

\begin{document}

%
%

\preprint{hep-ph/9703305}{PITHA 97/5}{}
\title{
       Next-to-Leading Order QCD Corrections to Three-Jet Cross Sections
       with Massive Quarks\footnote{supported  by BMBF,
                               contract 057AC9EP(1).}
        }
\author{
        Werner Bernreuther,
        Arnd Brandenburg\footnote{supported  by Deutsche
                                Forschungsgemeinschaft.},
        and Peter Uwer
        }
\address{
        Institut f\"ur Theoretische Physik, RWTH Aachen,
        D-52056 Aachen, Germany
        }
\date{\today}
\maketitle{}
\begin{abstract}
We calculate the cross section for
$e^+ e^-$ annihilation into three jets for massive quarks at
 next-to-leading order in perturbative QCD, both on and off the $Z$ resonance.
Our computation allows the implementation of any jet clustering algorithm.
We give results for the three-jet cross section involving $b$ quarks for the
JADE and Durham algorithm at c.m. energies $\sqrt s = m_Z$.
We also discuss a three-jet observable that is sensitive to
the mass of the $b$ quark.

\end{abstract}
\bigskip
\pacs{PACS number(s): 12.38.Bx, 13.87.Ce, 14.65.Fy }

\bigskip

Jets of hadrons, which originate from the
production and subsequent fragmentation
of quarks and gluons  in high energy electron positron annihilation
have been among the
key predictions \cite{EGR,SW} of
quantum chromodynamics (QCD). For precision tests of
QCD the $e^+e^-$ experiments provide a particularily clean set-up.
There exist a number of jet observables that are well-defined
(i.e., infrared-finite) in QCD, and which can be calculated
perturbatively as an expansion in the strong coupling $\alpha_s$.
 The next-to-leading order (NLO) QCD corrections to
three-jet production
were computed more than a decade ago \cite{ERT,FSKS} for massless quarks, and
subsequent implementations \cite{Lampe,KN,GG,FKS,CS} of  these results
 have been widely used for tests of QCD with
jet physics.

To date huge samples of jet events produced at the $Z$ resonance have been
collected both at LEP and SLC.
From these data  large numbers of jet events involving $b$ quarks can be
isolated with high purity using vertex detectors. For detailed
investigations of
$b$ jets quark mass effects must be taken into account in the
theoretical predictions
\cite{qmass}. Specifically, knowledge of  
the NLO three-jet fraction for non-zero quark mass opens
the possibility to measure
the mass of the $b$ quark from $b$ jet  data at the $Z$ peak \cite{BRS}.
Further applications include precision tests of the 
asymptotic freedom property
of QCD by means of three-jet rates measured at various
center-of-mass energies, also far below the $Z$ resonance \cite{Bethke}.

As far as massive quarks are concerned the three, four, and five jet rates
are known to leading order (LO) in $\alpha_s$ only \cite{Joffe,BMM,JLZ}.
In this Letter we report the calculation
of the $e^+ e^-$ annihilation cross section into three jets involving a massive
quark antiquark pair at
next-to-leading order QCD  \cite{next,Rod}.
The determination of this cross section  $\sigma_{NLO}^{3}$  
to order $\alpha_s^2$ consists of two parts:
First, the computation of the amplitude of the partonic reaction
$e^+e^-\rightarrow\gamma^\ast, Z^\ast\rightarrow Q \bar Q  g$
at leading and next-to-leading order in the  QCD coupling.  
Here $Q$ denotes a massive quark
and $g$ a gluon. We have calculated the complete decay amplitude and decay
distribution structure for this reaction. This allows for
predictions including
oriented three-jet events. The differential cross section involves the
so-called hadronic
tensor which contains five parity-conserving and four parity-violating
Lorentz structures.
Second, the leading order matrix elements of the
four-parton production processes
$ e^+e^- \rightarrow Z^\ast, \gamma^\ast \rightarrow
ggQ\bar{Q},Q\bar{Q}q\bar{q},Q\bar{Q}Q\bar{Q}$ are needed.
Here $q$ denote light quarks which are taken to be massless.
\par
The calculation of the ${\cal O}(\alpha_s^2)$
virtual corrections to the process
$e^+e^-\to Q\bar{Q}g$
with massive quarks is straightforward albeit tedious.
Non-neglection of the quark mass leads to a considerable
complication of the algebra. 
The infrared (IR) and ultraviolet (UV) singularities, which are encountered
in the computation of the one-loop integrals,
are treated within the framework
of dimensional regularization in $D=4-2\epsilon$ space-time dimensions.
We remove the UV singularities  
by the standard $\overline{\rm MS}$ renormalization.
An essential aspect of our computation is to show 
that the IR singularities of the virtual corrections are cancelled by the
singularities resulting from 
phase space integration of the squared tree amplitudes
for the production of four partons.
Different methods to perform  this cancellation have been developed
(see \cite{GG,FKS,CS} and references therein).
We use the so-called phase space slicing method elaborated in
\cite{GG}. The basic idea is to ``slice'' the phase space of the
four parton final state by introducing an unphysical parton
resolution parameter $s_{min}\ll sy_{cut}$, where $y_{cut}$ is
the jet resolution parameter. The parameter  $s_{min}$ splits the 
phase space into
a region where all four partons are ``resolved''
and a region where at least
one parton remains unresolved.
For massless partons, the resolved region may be conveniently
defined by the requirement that all invariants
$s_{ij}=(k_i+k_j)^2$ built from the parton momenta $k_i$ are
larger than the parameter
$s_{min}$. We have modified
this definition slightly to account for masses \cite{next}.
\par
In the unresolved region soft and collinear
divergences reside, which have to be isolated explicitly to cancel
the singularities of the virtual corrections.
This is considerably simplified due to collinear and soft
factorizations of the matrix elements
which hold in the limit $s_{min}\to 0$.
In the presence of massive quarks, the structure of
collinear and soft poles
is completely different as compared to the massless case. As an
example, we would like to discuss for
$e^+e^-\to Q(k_1)\bar{Q}(k_2) g(k_3)g(k_4)$
the limit where one gluon, say $g(k_4)$, becomes soft.
In this limit, the squared matrix element
can be written as a universal factor
multiplying the squared Born
matrix element for $e^+e^-\to Q\bar{Q} g$. The integration over
the soft gluon momentum $k_4$ can then be carried out analytically in
$D$ dimensions.
In the soft limit  $k_4\to 0$ the squared matrix
element reads
\bea
\label{eikonal}
|T_{fi}^{\scriptsize{\mbox{soft}}}(e^+e^-\to Q\bar{Q}gg)|^2
\ &=&\
\frac{g_s^2N_C}{2}\left[\sum_{a=1,2}\left(\frac{4t_{a3}}{t_{a4}t_{34}}-\frac
{4m^2}{t_{a4}^2}\right)
- \frac{1}{N_C^2}\left(\frac{4t_{12}}{t_{14}t_{24}}
-\frac{4m^2}{t_{14}^2}-\frac{4m^2}{t_{24}^2}\right)\right]\nonumber \\
\ &\times& \
|T_{fi}^{\scriptsize{\mbox{Born}}}(e^+e^-\to Q\bar{Q}g)|^2,
\eea
where $g_s$ is the strong coupling constant, $N_C=3$ is the number of colors,
 $t_{ij}=2k_ik_j$ and $m$ denotes the quark mass.
Each of the three terms in (\ref{eikonal}) can now be integrated
over the appropriate
soft phase space volume which we define by the conditions
$t_{a4}+t_{34} \ <\  2\, s_{min}$ (leading color terms), and
$t_{14}+t_{24} \ <\  2\, s_{min}$ (subleading color term).
The complete soft factor $S(k_1,k_2,k_3)$ multiplying the
squared Born matrix element for $e^+e^-\to Q(k_1)\bar{Q}(k_2)g(k_3)$
which is obtained by this integration reads:

\bea
\label{softfactor}
S(k_1,k_2,k_3)\ &=&\ \frac{\alpha_s}{4\pi}\,N_C\,\,\frac{1}
    {\Gamma(1-\epsilon)}\,
    \left(\frac{4\pi\mu^2}{s_{min}}\right)^{\epsilon}
\Bigg[
    \Bigg\{
        \left(\frac{s_{min}}{t_{13}+m^2}\right)^{-\epsilon}
       \Bigg( \frac{1}{\epsilon^2}
        -  \frac{1}{\epsilon}
        \bigg[\ln\left(1+\frac{t_{13}}{m^2}\right)
 \nonumber \\ \ &+& 2\,\ln(2) - 1 \bigg]
     - \frac{\pi^2}{6}+2\ln^2(2)-2\ln(2)
        +\Big[2\ln (2)+\frac{2m^2}{t_{13}}+1\Big]\,
        \ln\left(1+\frac{t_{13}}{m^2}\right) \nonumber \\ \ &-& \
        \frac{1}{2}\ln^2\left(1+\frac{t_{13}}{m^2}\right)
      - 2\,\Li{\frac{t_{13}}{t_{13}+m^2}}
        \Bigg) + (t_{13}\leftrightarrow t_{23})
     \Bigg\} \nonumber \\ \ &-&\ \frac{1}{N_C^2}
    \left(\frac{s_{min}}{t_{12}+2m^2}\right)^{-\epsilon} \frac{1}{\beta}
    \Bigg(\frac{1}{\epsilon}\left[\,2\beta+ (1+\beta^2)\ln(\omega)\right]
     - 4\beta\ln(2) - 2\ln(\omega)
    \nonumber \\
    \ &-&\ 2 \ln(2) (1+\beta^2)\ln(\omega)
    -\frac{1+\beta^2}{2}\ln^2(\omega)-
    2(1+\beta^2)\Li{1-\omega} \Bigg)\Bigg]
    + {\cal O}(\epsilon) ,
\eea

\noindent where $\beta=\sqrt{1-4m^2/(t_{12}+2m^2)}$,
$\omega=(1-\beta)/(1+\beta)$,
and $\mu$ is an arbitrary
scale introduced to keep $\alpha_s$ dimensionless in $D$ dimensions.
The poles in $\epsilon$
exhibited in (\ref{softfactor}) (and additional poles from the
collinear region of phase space which we do not show explicitly) 
can now be cancelled against the IR poles of the
one-loop integrals entering the virtual corrections.
One is then left with a completely regular 
differential three-parton cross
section which depends on $s_{min}$.
\par
The contribution to $\sigma_{NLO}^3$ of 
the ``resolved'' part of the four-parton cross section is finite
and may be evaluated in $D=4$ dimensions, which is
of great practical importance.
It also  depends on $s_{min}$ and is  most conveniently obtained 
by a numerical integration.
Since the parameter $s_{min}$ is
completely arbitrary, the sum of all contributions 
to $\sigma^3_{NLO}(y_{cut})$ must not depend on $s_{min}$.
In the soft and collinear approximations 
one neglects terms which vanish
as $s_{min}\to 0$. This limit can  be carried out numerically.
Since the individual
contributions depend logarithmically
on $s_{min}$, it is a nontrivial test of the
calculation to demonstrate that
$\sigma_{NLO}^3$ becomes independent of
$s_{min}$ for small values
of this parameter. Moreover, in order to avoid large numerical
cancellations, one should determine the largest value of $s_{min}$
which has this property.
\par
The  three jet cross section depends on the
experimental jet definition.
We consider here the JADE \cite{Jade} and Durham \cite{Dur}
clustering algorithms, although other schemes \cite{BKSS} can also be
easily implemented. We have checked that 
we recover the result of \cite{KN} in the massless limit.
\par
We now present our results for
the cross section $\sigma_{NLO}^{3,b}$ for $b$ quarks in the 
$\overline{\rm MS}$  scheme. We require that at least two 
of the jets that remain after the 
clustering procedure contain a $b$ or $\bar{b}$ quark \cite{qqbb}.  
We use the $b$ quark mass parameter  $m_b(\mu)$
defined in the  $\overline{\rm MS}$ scheme at a scale $\mu$.
The asymptotic freedom property of QCD predicts that
this mass parameter decreases when being evaluated at a higher scale.
With $m_b(m_b)$ = 4.36 GeV \cite{Neubert} 
and $\alpha_s(m_Z)$ = 0.118 \cite{PDG}
as an input and employing the
standard renormalization group evolution of the coupling and
the quark masses, we use the value
$m_b(\mu=m_Z)$ = 3 GeV. 
\par
Figs. 1a,b show the three jet cross section
$\sigma_{NLO}^{3,b}$ at $\sqrt{s}=\mu=m_Z$ with $b$ quarks of mass
$m_b= 3$ GeV as a function of $y_{min}=s_{min}/(sy_{cut})$
for the JADE and Durham
algorithms at a value of the jet resolution
parameter $y_{cut}=0.03$.
It can be clearly seen that
the cross section reaches a plateau for small values of
the parameter $y_{min}$. The error of the numerical
integration becomes bigger as $y_{min}\to 0$. In order to keep
this error as small as possible without introducing
a systematic error from using
the soft and collinear approximations,
we take $y_{min}=10^{-2}$ for the JADE algorithm and
$y_{min}=5\times 10^{-3}$ for the Durham algorithm.
These values are used in Figs. 1c,d, where we plot 
$\sigma_{NLO}^{3,b}$ as a function of $y_{cut}$
together with the LO result. 
The QCD corrections to the LO result are quite sizable,
as known also in the massless case.
The renormalization scale dependence which is also shown in
Figs. 1c,d is modest in the whole $y_{cut}$ range exhibited for the Durham
and above $y_{cut}\sim 0.01$ for the JADE
algorithm. Below this value perturbation theory is
not applicable in the JADE scheme.    
\par
The effects of the $b$ quark mass at the $Z$ peak may be exhibited
with the following ratio \cite{delphi}
\begin{equation}\label{obs}
{\cal B}(y_{cut})=\frac{R_3^b(y_{cut})}{R_3^{udsc}(y_{cut})}.
\end{equation}
Here we define $R_3^b=\sigma^{3,b}/\sigma(e^+e^-\to b\bar{b})$, 
and 
$R_3^{udsc}$ is the three-jet fraction for the four light quarks
with no flavor tagging. (Note that ${\cal B} \neq 1$ in the limit
 $m_b\to 0$ due to the different definitions of  $R_3^b$ and 
$R_3^{udsc}$.) 
The LO and NLO results for the observable ${\cal B}$ 
are shown in Figs. 2a,b for the JADE and the Durham algorithms. 
We took the massless ${\cal O}(\alpha_s^2)$
results from \cite{KN,BKSS}. 
As both the LO and NLO terms in $R^b_3$ depend on $m_b$
it is clear that comparison of
our result with measured values of ${\cal B}(y_{cut})$ 
would allow for an
unambiguous determination of
$m_b$ within a given renormalization scheme.
In view that the $b$ mass effect in ${\cal B}$
 -- and in other observables, for 
instance the differential two-jet
distribution \cite{next} -- is only 
of the order of a few percent this constitutes
an experimental challenge. Moreover, further work is 
needed to assess in detail
the theoretical uncertainties involved \cite{tag}.
Yet such  an analysis would be worth the effort: 
it would be the first
determination of the $b$ quark mass at a high scale, 
and it might also
experimentally establish the ``running'' of a quark mass 
as predicted by QCD.
\par
Summarizing we have computed the NLO QCD corrections 
for $e^+e^-\to 3$
jets for massive quarks. Our results, which we shall report on
in detail in future work, should find 
applications to a number
of precision tests of QCD involving $b$ and $c$ quarks at various
c.m. energies, and
to theoretical investigations 
of top quark production
at very high-energetic electron positron collisions.
\bigskip

\section*{Acknowledgments}
We thank S. Bethke, O. Biebel, P. N. Burrows, 
P. Haberl, G. Rodrigo, and M. Wunsch for discussions.
\bigskip

\newpage
\unitlength 1cm
\begin{picture}(15,5)
\put(1.25,0){\psfig{figure=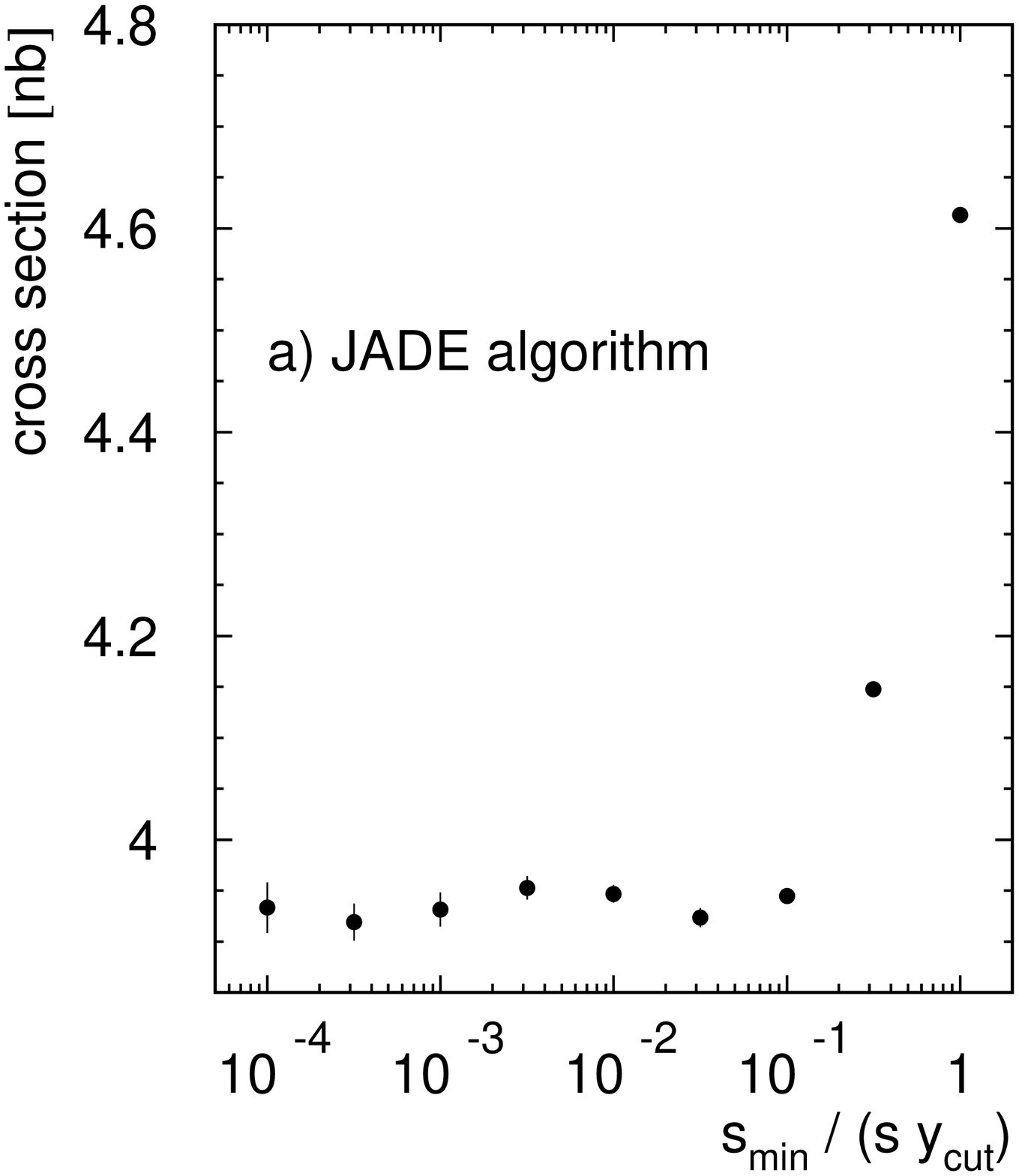,height=6.05cm,width=5.5cm}}
\put(8.,0){\psfig{figure=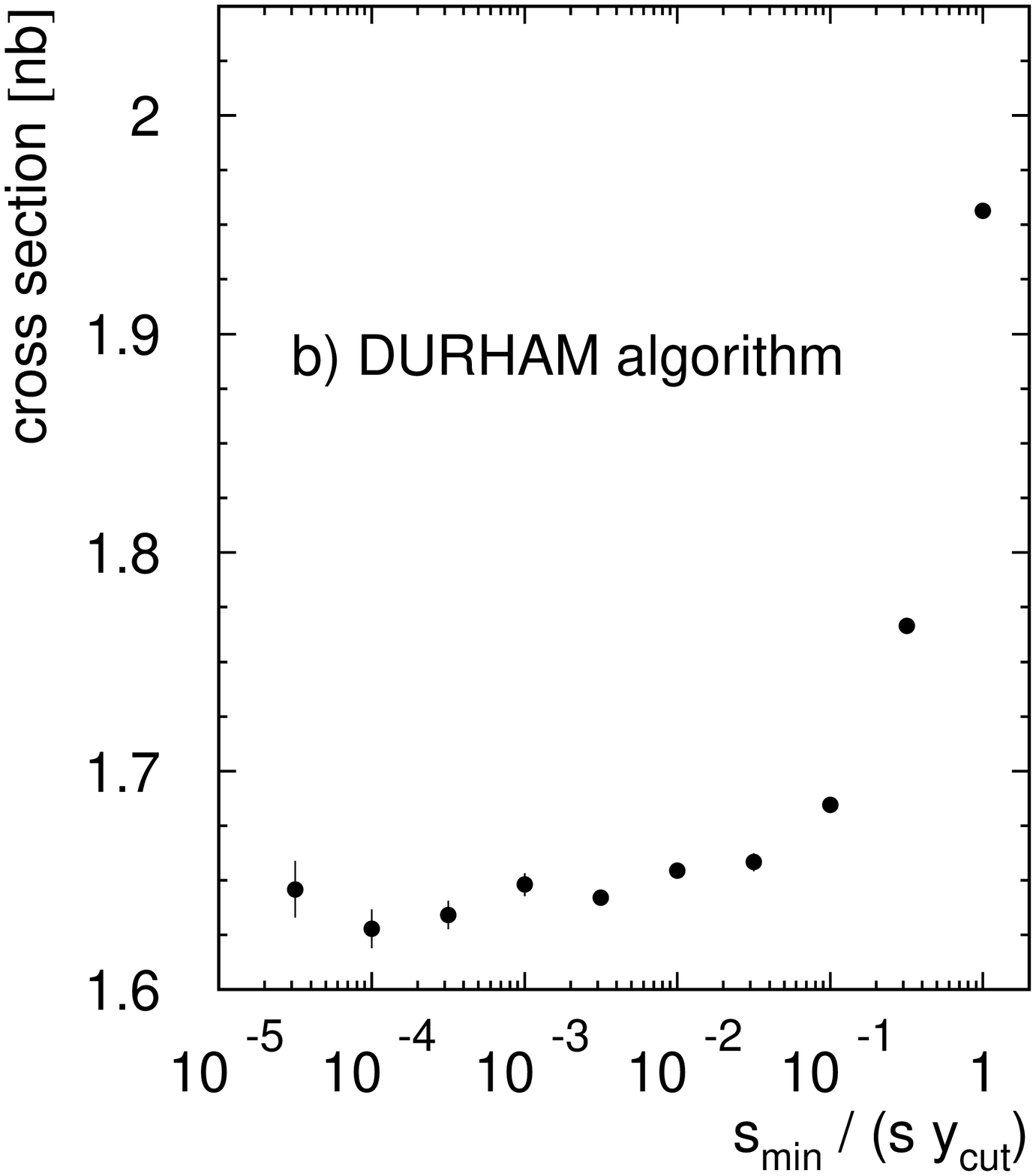,height=6.05cm,width=5.5cm}}
\put(1.25,-6){\psfig{figure=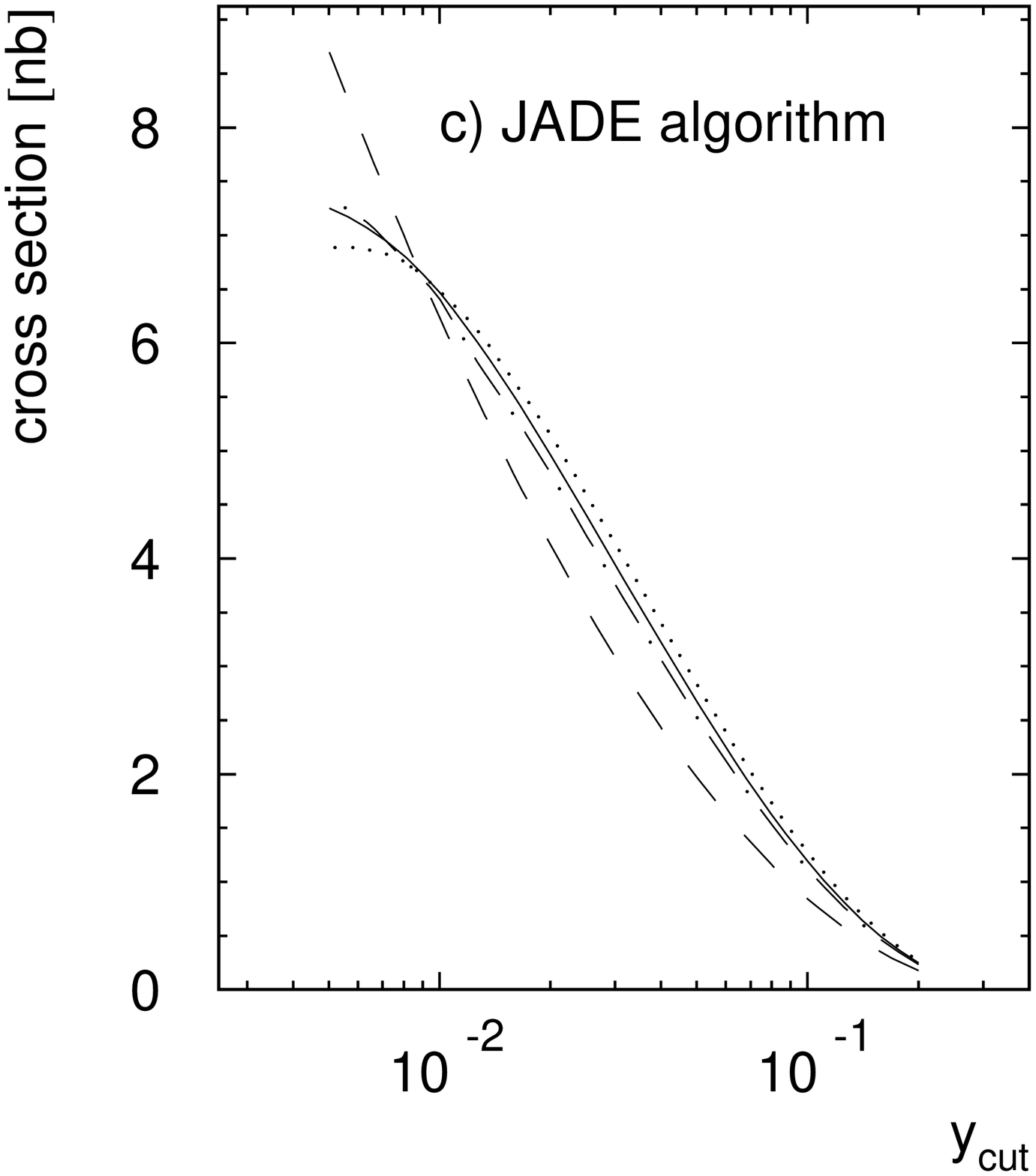,height=6.05cm,width=5.5cm}}
\put(8.,-6){\psfig{figure=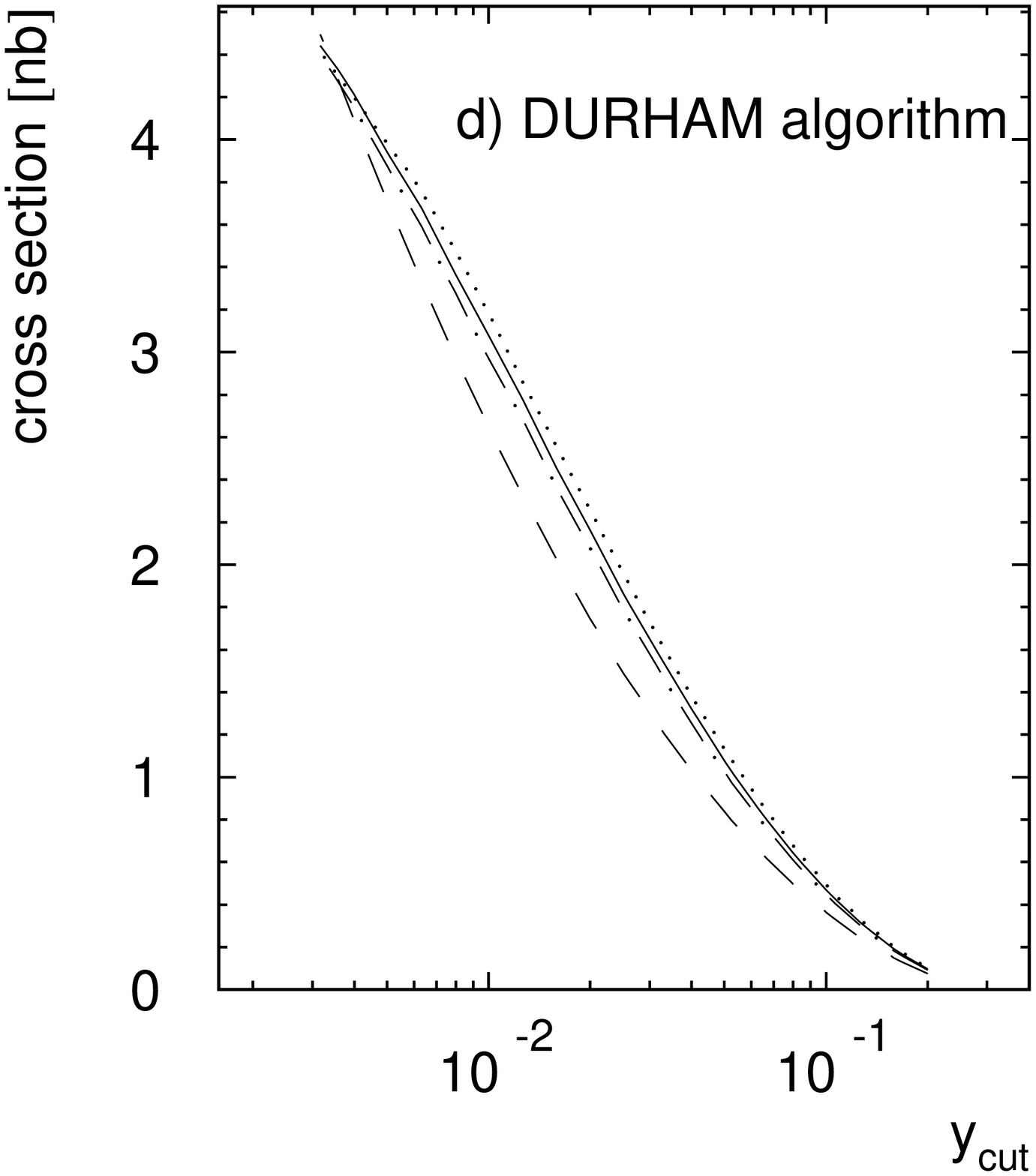,height=6.05cm,width=5.5cm}}
\put(0,-6.5)
{\begin{minipage}[t]{15.5cm}
\baselineskip14pt
\small {\bf{Figure 1:}}
Figs. 1a and 1b show 
$\sigma_{NLO}^{3,b}$ as defined in the text at 
$\sqrt{s}=\mu=m_Z$ as a function of $y_{min}=s_{min}/(sy_{cut})$
for the JADE and Durham algorithms at a value of the jet resolution
parameter $y_{cut}=0.03$ with $m_b(\mu=m_Z)=3$ GeV and 
$\alpha_s(\mu=m_Z)=0.118$.
Figs. 1c and 1d show $\sigma^{3,b}$ 
as a function of $y_{cut}$ for the JADE and Durham algorithms,
respectively.
The dashed line is the LO result.
The NLO results are for $\mu=m_Z$ (solid line),
$\mu=m_Z/2$ (dotted line), and $\mu=2m_Z$ (dash-dotted
line). Initial-state photon radiation is not included in the cross
sections.
\end{minipage}}
\end{picture}
\newpage
\unitlength 1cm
\begin{picture}(15,5)
\put(0.8,0){\psfig{figure=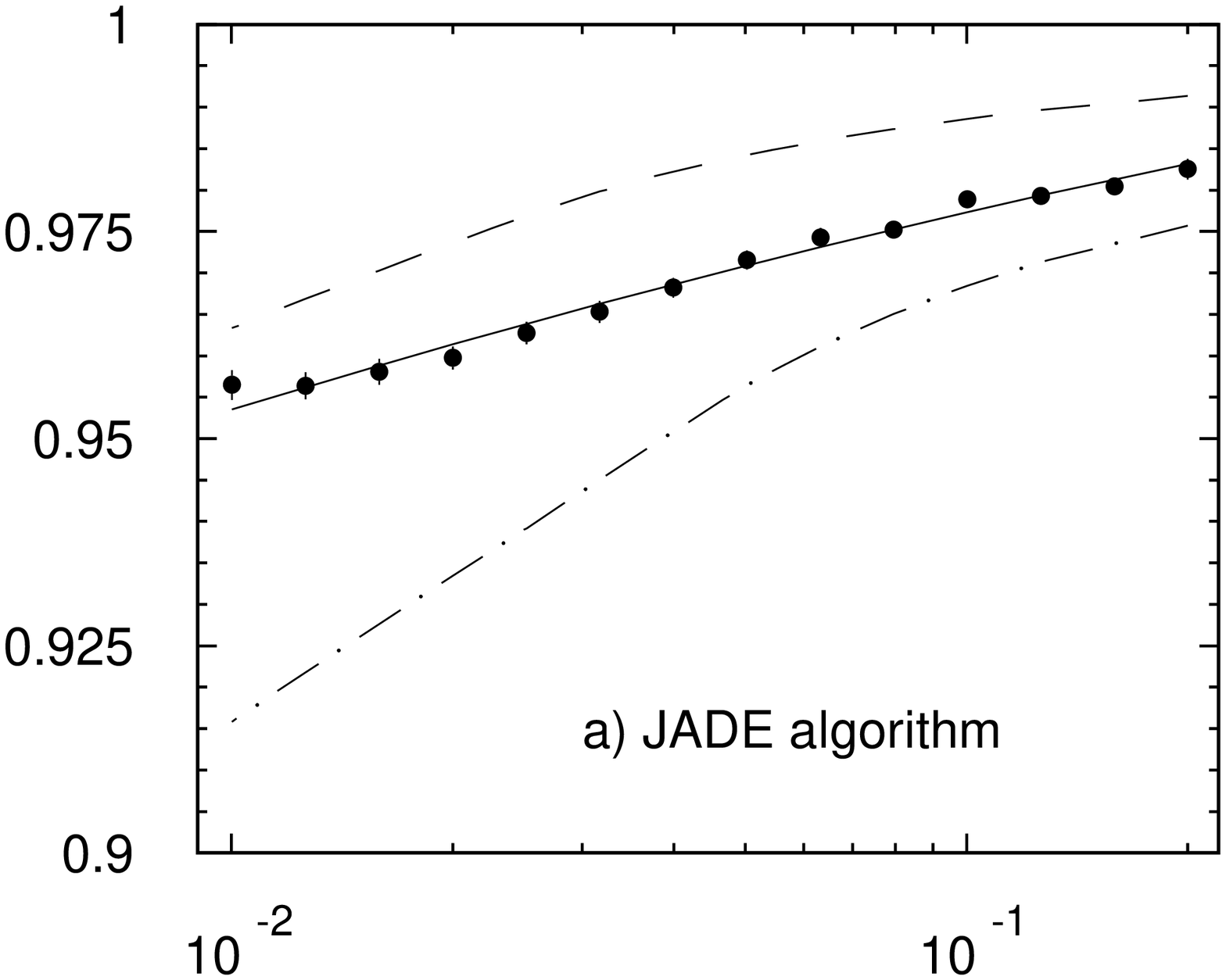,height=5.4cm,width=6.6cm}}
\put(7.55,0){\psfig{figure=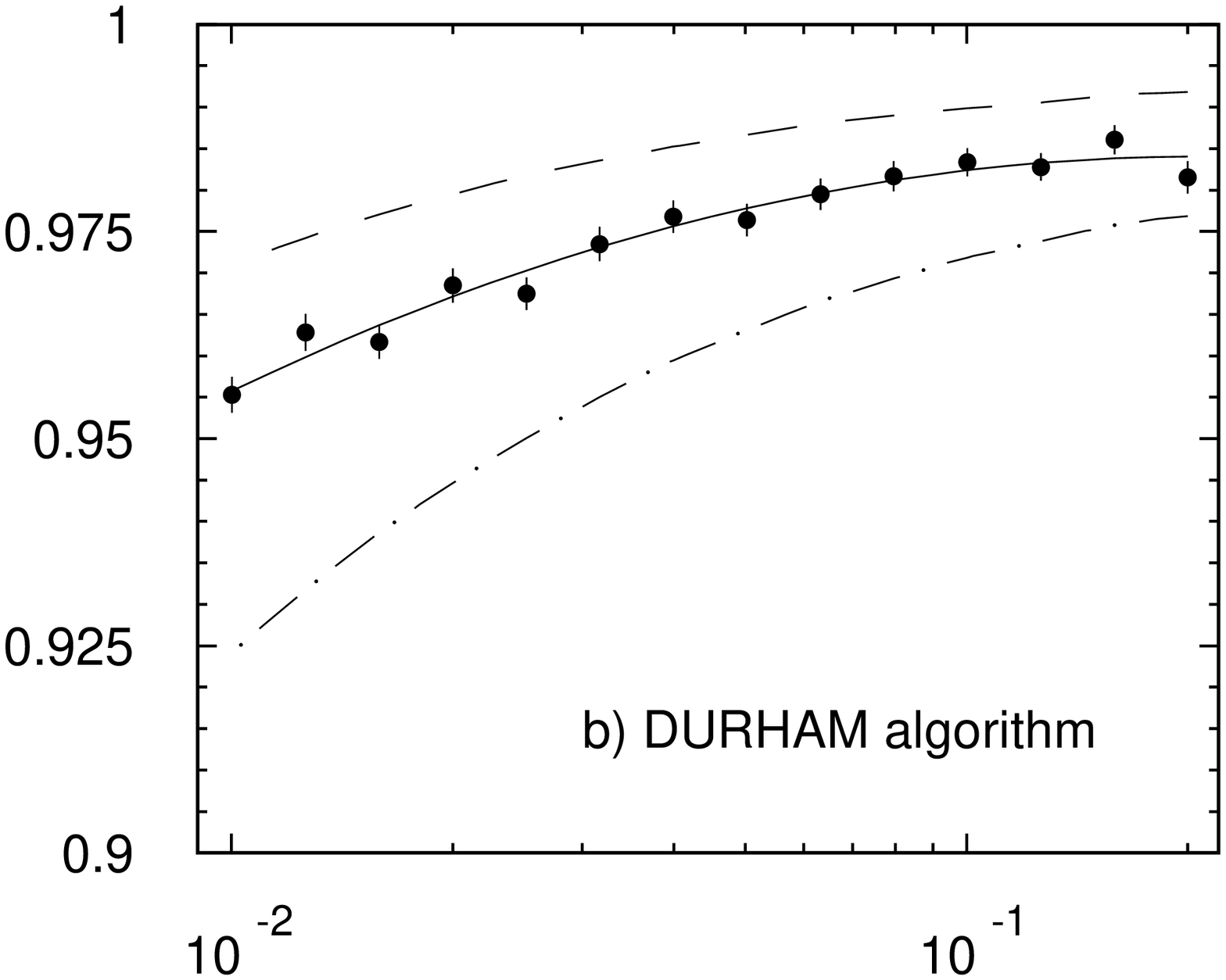,height=5.4cm,width=6.6cm}}
\put(0,-0.5)
{\begin{minipage}[t]{15.5cm}
\baselineskip14pt
\small {\bf Figure 2:}
The ratio ${\cal B}$ of eq. (\ref{obs}) as a 
function of $y_{cut}$ for (a) the JADE  and (b) the Durham algorithm,
using values for $m_b$ and $\alpha_s$ as in Fig. 1.
The dashed line is the LO result. The points and the solid line
are the NLO result  for $\mu=m_Z$.
For illustrative purposes, the dash-dotted line 
shows the LO result for $m_b=5$ GeV.
\end{minipage}}
\end{picture}


\begin{thebibliography}{99}
\bibitem{qmass}
While the mass of the $b$ quark can be neglected to good approximation
in the computation of the total $b\bar{b}$ production rate at the $Z$ peak, the
n-jet cross sections depend, apart from the c.m. energy, on an extra scale,
to wit, the jet resolution parameter $y_{cut}$. For small $y_{cut}$ 
effects due to the $b$ quark mass can be enhanced.
\bibitem{next}
Details of our calculation will be published elsewhere.
\bibitem{qqbb} We  include also the contribution from the diagrams
$e^+e^-\to q\bar{q}g^*\to q\bar{q}b\bar{b}$. Requiring
two $b$ quark jets leads to an infrared-safe 
three-jet cross section in the
limit $m_b\to 0$ since it forbids the 
clustering of the $b\bar{b}$ pair from $g^*$ into
a single jet, which would lead to large logarithms 
involving $m_b$. If one would
allow such a clustering, the logarithms should be 
absorbed into a fragmentation
function for a gluon into a B hadron.
\end{thebibliography}
\end{document}